\documentstyle[aps,prl]{revtex}
\begin{document}

\title{Long-range interacting solitons: pattern formation and nonextensive thermostatistics}
\author{L. E. Guerrero\thanks{Corresponding author. Fax: +582-9063601; e-mail: lguerre@usb.ve}}
\address{Departamento de F\'{\i}sica, Universidad Sim\'on Bol\'{\i}var, Apartado Postal 89000,
Caracas 1080-A, Venezuela}
\author{J. A. Gonz\'alez}
\address{Centro de F\'{\i}sica, Instituto Venezolano de Investigaciones
Cient\'{\i}ficas, Apartado Postal 21827, Caracas 1020-A, Venezuela}
\date{\today}
\maketitle
\begin{abstract}
The nonlinear Klein-Gordon equation with a different potential that
satisfies the degeneracy properties discussed in this paper possesses
solitonic solutions that interact with long-range forces. We generalize the Ginzburg-Landau equation in such a way that the
topological defects supported by this equation present long-range
interaction both in $D=1$ and $D>1$. Finally, we construct a system of two equations with two complex order parameters for which the interaction forces between the
topological defects decay so slowly that the system enters the
nonextensivity regime.

\end{abstract}

\section{Introduction}

There is a great interest in the formulation of models in which the solitons
can interact with long-range forces\cite{r:1,r:2,r:3,r:4,r:5,r:7}. The
soliton solutions of the well-known models (e.g., sine-Gordon and $\phi ^4$
equations) interact with short-range forces\cite{r:1}. There is experimental
evidence\cite{r:4,r:5} that most real transfer mechanisms have long-range
character. We are interested in models where there is spontaneous formation
of particle-like objects that possess long-range interactions. In such
systems we can study pattern formation and other complex phenomena. Due to
the fact that systems with long-range interactions can exhibit nonextensive
behavior\cite{r:8,r:9,r:10}, the models we are investigating are relevant
for the recently proposed thermostatistical theories\cite{r:8}.

Some authors have considered long-range effects\cite{r:4} including ad-hoc
nonlocal terms in the equations. Spin systems have been also studied
considering the coupling constant $J_{ij}$ between the lattice spins to be
proportional to $r_{ij}^{-\alpha }$\cite{r:10}. Gonz\'alez and
Estrada-Sarlabous demonstrated\cite{r:1} that pure Klein-Gordon equations without
coordinate-dependent terms, can support solitons with long-range
interactions.

In this paper we investigate a system that is an extension of the
Klein-Gordon equation and possesses soliton solutions with long-range
interaction. We also introduce a generalized version of the Ginzburg-Landau
equation which supports topological defects whose interaction force decays
very slowly. It is possible to create a gas of such topological defects with
an interaction force that decays so slowly that we enter the nonextensivity
regime. We apply these results to nonequilibrium systems, pattern formation
and growth models.

\section{Modified Klein-Gordon equation}

In this section we will study the Klein-Gordon equation 
\begin{equation}
\phi _{tt}-\phi _{xx}-G(\phi )=0.  \label{eq:1}
\end{equation}

Here $G(\phi )=-\frac{\partial U(\phi )}{\partial \phi }$. In Ref. \cite
{r:1,r:11} the authors investigated systems of type (\ref{eq:1}) where $%
U(\phi )$ possesses at least two minima (at points $\phi _1$ and $\phi _3$)
and a maximum at point $\phi _2$ ($\phi _1<\phi _2<\phi _3$). In particular,
it was shown that if the potential $U(\phi )$ behaves in the neighborhood of
a minima as $U(\phi )\sim \left( \phi -\phi _i\right) ^{2n}$, then the
solitons supported by the system interact with a force $F(d)$ that decays
exponentially with the distance for $n=1$. On the other hand, for $n>1$ the
solitons interact with a force that decays following the law $F\sim d^{\frac{%
2n}{1-n}}$.

\subsection{Pattern formation}

The most investigated growth model is the KPZ equation\cite{r:15}.
Nevertheless other models have been proposed\cite{r:16} including the
sine-Gordon model\cite{r:17,r:18}.

In this section we present an alternative model which is given by the
equation

\begin{equation}
\phi _{tt}+\gamma \phi _t-\phi _{xx}-G(\phi )=\eta (x,t),  \label{eq:15}
\end{equation}

where $\eta (x,t)$ is spatiotemporal white noise with the properties $%
\left\langle \eta (x,t)\right\rangle =0$, $\left\langle \eta (x,t)\eta
(x^{\prime },t^{\prime })\right\rangle =2D\delta (t-t^{\prime })\delta
(x-x^{\prime })$. Here the potential is $U(\phi )=2\sin ^{2n}\left( \frac
\phi 2\right) $. It can be shown that this potential has the property $%
U(\phi )\sim \left( \phi -\phi _i\right) ^{2n}$.This equation with $n>1$ can
be used as a growth model for periodic media with marginal stability.

The growth model described by equation (\ref{eq:15}) presents noise-induced
pattern formation that can be studied through the calculation of the
roughness exponent. The sine-Gordon equation ($n=1$) does not eliminate
disorder at great scales\cite{r:18}. Meanwhile, the systems with $n\gg 1$
display self-affine behavior at all scales\cite{r:7}. The long-range
interaction between the solitons is the most relevant feature of this
system. This also can explain the wavelet analysis performed in Ref. \cite
{r:7}. There for $n\gg 1$ the authors found the existence of structures at
all scales.

Summarizing, in this system there is pattern formation produced by the
spontaneous generation of topological objects with long-range interactions
which can create complex structures exhibiting fractal behavior.

\section{A generalized Ginzburg-Landau equation}

In this section we introduce a generalized version of the Ginzburg-Landau
equation:

\begin{equation}
\frac{\partial u}{\partial t}=\nabla ^2u+u(1-\left| u\right| ^2)^{2n-1}.
\label{eq:6}
\end{equation}

Note that for $n=1$ we recover the well-known Ginzburg-Landau (G-L) equation%
\cite{r:12,r:13}. Equation (\ref{eq:6}) preserves all the qualitative
properties of the G-L equation even for $n>1$. There exist an unstable state
at $u=0$ and a degenerate stable state at $\left| u\right| =1$. For all
integer $n$, the Equation (\ref{eq:6}) possesses topological solitons.
However, for $D=1$ ($n=1$) G-L solitons interact with a force that decays
exponentially. On the other hand the soliton solutions of Equation (\ref
{eq:6}) ($n>1$) interact with long-range forces as the modified sine-Gordon
equation.

Nevertheless, here we will be interested in vortice-like topological defects
in $D=2$. A vortice-like topological defect with topological charge $\kappa $
can be expressed in polar coordinates ($r$, $\varphi $) by the following
equation $u=\rho (r)e^{i\kappa \varphi }$, where $\rho (r)$ is a solution of
the equation

\begin{equation}
\frac{\partial ^2\rho }{\partial r^2}+\frac 1r\frac{\partial \rho }{\partial
r}-\frac{\kappa ^2\rho }{r^2}+\rho \left( 1-\rho ^2\right) ^{2n-1}=0.
\label{eq:8}
\end{equation}

The analysis of the asymptotic behavior of the vortice solution for $n=1$ of
Equation (\ref{eq:8}) yields that in the limit $r\rightarrow \infty $, $%
\left( \rho (r)-1\right) \sim r^{-2}$. If $n>1$, then $\left( \rho
(r)-1\right) \sim r^{\frac 1{1-n}}$. Note that the transition from
$n=1$ to $n>1$ is not continuous because for the long-range
potentials that we are considering\cite{r:1,r:7,r:14} $n$ is
usually taken as an integer number. In principle, $n$ can be
generalized to be a real number. In this case, we define
$n=1+\varepsilon $. Hence for $\varepsilon <\frac 12$ we have  
$\left( \rho (r)-1\right) \sim r^{-2}$. On the other hand for
$\varepsilon >\frac 12$, 
$\left( \rho (r)-1\right) \sim r^{-\frac 1\varepsilon }$. As
we can see the solutions for $\varepsilon <\frac 12$
and $\varepsilon >\frac 12$ 
match if $\varepsilon$ is real.

In the case of the generalized G-L equation (\ref{eq:6}) the force that acts
on a vortice situated at the point $r$ due to the existence of another
vortice at the coordinates origin satisfies the relation $F \sim \left( \rho
(r)-1\right)$. Thus the vortices produced by equation (\ref{eq:6}) with $n=1$
have Coulomb interactions. Meanwhile, for $n>1$ the interaction decays much
more slowly.

\section{Nonextensivity}

In principle, it is possible to construct a system described by equations of
the type

\begin{equation}
\frac{\partial ^2\phi _1}{\partial t^2}+\gamma \frac{\partial \phi _1}{%
\partial t}-\nabla ^2\phi _1=-\frac{\partial V\left( \left| \phi _1\right|
,\left| \phi _2\right| \right) }{\partial \phi _1},  \label{eq:11}
\end{equation}

\begin{equation}
\frac{\partial ^2\phi _2}{\partial t^2}+\gamma \frac{\partial \phi _2}{%
\partial t}-\nabla ^2\phi _2=-\frac{\partial V\left( \left| \phi _1\right|
,\left| \phi _2\right| \right) }{\partial \phi _{2}},  \label{eq:12}
\end{equation}

where the potential $V\left( \left| \phi _1\right| ,\left| \phi _2\right|
\right) $ holds the necessary conditions in order to produce long-range
interactions. When we are in the presence of systems of equations like (\ref
{eq:11},\ref{eq:12}) with two order-parameters we can have the situation
where the sustained topological defects repel each other at very small
distances and they attract each other at great distances\cite{r:14}. We can
have an effective interaction potential like the following:

\begin{equation}
V_{eff}(r)=\varepsilon \left[ \left( \frac \sigma {1+r^2}\right) ^{\rho
/2}-\left( \frac \sigma {1+r^2}\right) ^{\alpha /2}\right] ,  \label{eq:13}
\end{equation}

where $\alpha <\rho $. This is a situation equivalent to that discussed in
Ref. \cite{r:9}. Thus when we have $N$ particles in the system, the energy
will grow with $N$ following the laws:

\begin{eqnarray}
E\sim \cases{N &if $\frac \alpha D>1$, \cr N\ln N &if $\frac \alpha D=1$,
\cr N^{2-\frac \alpha D} &if $\frac \alpha D<1$. }  \label{eq:14}
\end{eqnarray}

In our case $\alpha =\frac{2-n}{n-1}$. When $n>1$ (for $n$ integer and $D=2$)
we are in the nonextensive regime. In the general case, for $n=1+\varepsilon
$ (real) we obtain the nonextensivity condition $\frac{1-\varepsilon }%
\varepsilon <D$.

This work has been partially supported by Consejo Nacional de
Investigaciones Cient\'{i}ficas y Tecnol\'ogicas (CONICIT) under Project
S1-2708.

\end{document}